\documentclass[mathleft
]{an}
\usepackage{graphicx}
\usepackage{times}
\usepackage{natbib}
\bibpunct{(}{)}{;}{a}{}{,}

\begin{document}

\Pagespan{789}{}
\Yearpublication{2006}%
\Yearsubmission{2005}%
\Month{11}%
\Volume{999}%
\Issue{88}%

\title{Samples and statistics of CSS and GPS sources}

\author{M. Giroletti\inst{1}\fnmsep\thanks{Corresponding author:
  \email{giroletti@ira.inaf.it}\newline}
\and
A. Polatidis\inst{2}
}
\titlerunning{Samples/Statistics}
\authorrunning{M. Giroletti \& A. Polatidis}
\institute{
INAF Istituto di Radioastronomia
via Gobetti 101, 40129 Bologna, Italy
\and
Joint Institute for VLBI in Europe (JIVE), Postbus 2, 7990 AA, Dwingeloo, The Netherlands}

\received{}
\accepted{}
\publonline{later}


\abstract{%
  Several samples have been proposed in the last years in order to study the
  properties of intrinsically small sources. In this paper, we review the
  properties of the main samples that are currently available, both selected on
  the basis of spectral index and of morphology. As a result of the work in
  this area, large numbers of intrinsically small sources have been found. We
  summarize the present status of hot spot advance measurements, listing 18
  sources with available VLBI data. The mean hot spot separation velocity is
  $v_\mathrm{sep} = (0.19 \pm 0.11)h^{-1}c$ and the kinematic ages span the
  range from 20 to 3000 years. Finally, we present a brief outlook on the use
  of future instrumentation in order to improve our understanding of radio
  source evolution. Prospects for VSOP2, e-VLA, e-MERLIN, LOFAR, ALMA, and
  Fermi are suggested.  }

\maketitle

\section{Introduction}

Studies of individual objects are certainly valuable to gain information about
the physics of radio sources. However, a general understanding of the
triggering and evolution of radio activity can only be obtained with a
population-based approach. Of course, since one can not study {\it all}
galaxies, we are forced to select samples that we believe are best suited to
answer the basic questions we are facing.  Since all selection criteria are
somehow biasing our results, it may be important to summarize the
characteristics of the various works. This is particularly relevant in our
field, where selection on spectra and morphology can be really confusing!

The first studies on Compact Steep Spectrum (CSS) and Gigahertz Peaked
Spectrum (GPS) sources were performed on the 3CR and the Peacock \& Wall
catalogues \citep{Spencer1989,Fanti1990}. The number of sources selected in the
early days did not exceed a few dozens, and these sources were typically quite
powerful. By the time of the Kerastari meeting (2002), the interest in larger
and deeper samples was already clear, as documented by the appearance of works
based on the B3/VLA sample \citep[$S_{0.4} > 0.8$ Jy,][]{Fanti2001}, the S4
sample \citep[$S_5>0.5$ Jy,][]{Saikia2001}, and the FIRST survey
\citep[$S_5>0.15$ Jy,][]{Marecki2003}. In the concluding remarks, the need for
additional samples of radio sources was clearly stated, in order to inform us
about their volume densities and evolution \citep{Woltjer2003}.

However, building a sample of young sources is a task that takes a lot of
effort. First, since our interest lies in the short initial part of the life of
radio sources, we are by definition looking for rare targets. Therefore, large
areas have to be investigated, requiring large fields of view but without
giving up on resolution.

Secondly, the well known relation between peak frequency ($\nu_\mathrm{peak}$)
and linear size ($LS$) and the large possible range of $\nu_\mathrm{peak}$
(between about 100 MHz and some 10's GHz) make it extremely inefficient to
select candidates of all sizes/ages. As a consequence, samples tend to focus on
sub-classes and it is difficult to obtain an unbiased and homogeneous view of
the evolution.

Finally, the follow up work can also take a considerable amount of
time. Optical data are needed to determine the host type (galaxy vs.\ QSO) and
possibly the redshift (to tie the angular dimension to a linear scale). VLBI
images are also important to exclude core-jet sources and eventually determine
the kinematic ages, a process that by itself requires a few to several years.

In the remainder of this paper, we will review some of the main results
obtained on the study of samples of CSS and GPS sources in the period between
the Kerastari meeting (2002) and today. Section \ref{s.samples} will be devoted
to a survey of the various samples, divided among those selected on the
spectral properties and on the morphology, with a short discussion of the two
approaches. In Sect.~\ref{s.stats}, we will report some statistics and in
particular we will summarize the present status on hot spot advance
measurements. Finally, some suggestions for the future are given in
Sect.~\ref{s.future}.

\section{Samples} \label{s.samples}

\subsection{Spectrally selected samples}

\subsubsection{The B3/VLA CSS Sample}

Starting from the 1049 sources in the B3/VLA sample \citep{Vigotti1989},
\citet{Fanti2001} have searched for Compact Steep Spectrum sources, imposing
selection criteria on both compactness ($LS< 20 h^{-1}$ kpc) and spectral index
($\alpha_{1.5}^{10} > 0.5$). A further selection was applied on the low
frequency flux density ($S_{0.4\,\mathrm{GHz}} > 0.8$ Jy), providing a final
sample of 87 sources.

This number makes it the largest sample of CSS sources with high quality
information; statistically, the 87 sources represent the 23\% of the population
in the corresponding flux density bin, a percentage similar to that found in
earlier works by \citet{Fanti1995}.  Thanks to optical spectra or photometry,
62\% of the sources have a measured redshift; the final range is $0.02<z<4.1$,
with a median $z=1.05$. This quite large distance range combined with the flux
density measurements corresponds to radio powers in the range
$10^{25}<P_{0.4\,\mathrm{GHz}}<10^{29}$ [W Hz$^{-1}$]. Therefore, besides a
number of very luminous sources, some members of the sample are also relatively
low power radio sources.

The first paper, beside defining the sample, presented also the results of
multi-frequency VLA observations and a discussion of properties such as the
morphology, the presence of spectral steepening at high frequency, and the
detection of significant polarization (4\% at 5 GHz, and 6\% at 8
GHz). Moreover, a study of the number counts with respect to the largest linear
size ($LLS$), revealed a power law relation $dN/d(LLS) \propto LLS^{-0.6}$.

Follow up studies on the sample have greatly clarified the nature of both the
sources and the medium in which they advance. Multi frequency observations with
various interferometers (EVN, VLBA, MERLIN) have probed different angular
scales, revealing both Compact Symmetric Objects (CSO) and Medium Symmetric
Objects (MSO), besides a number of naked cores
\citep{Dallacasa2002a,Dallacasa2002b,Orienti2004}. High-frequency,
high-resolution data have also been obtained with the VLA at 15 GHz. This has
also permitted a spectral index study over a large range of frequency; the
continuous injection model is in good agreement with the data for $\ge85\%$ of
the sources \citep{Rossetti2006}. Finally, polarization observations have also
been obtained, and they have been exploited to discuss the ambient Faraday
medium properties \citep{Fanti2004,Rossetti2008,Rossetti2009}.

In total, the sample has produced a series of 7 papers (so far) and a large
number of citations (90 at the time of writing). Thanks to its large size and
the huge amount of work with a broad range of techniques, it is probably the
best source of knowledge about not only the evolution of radio sources but also
the properties of the inner regions of galaxies. It would be great if the
B3/VLA sample were expanded to a lower flux density threshold, although it will
be difficult to keep up with the great quality that has been done so far.

\subsubsection{FIRST-based survey(s) of CSS sources}

Another sample that has given rise to a large number of works of high quality
and impact is the one developed by the Toru\`n group in the same years. It was
discussed at the Kerastari meeting \citep{Marecki2003}, although a more
detailed discussion is surely the one given in \citet{Kunert2002}. The starting
ground for this sample is the FIRST survey, which was supposed to provide
information about structure. Indeed, compactness ($LAS < 3\arcsec$) and steep
spectrum ($\alpha_{1.6}^5 > 0.5$) are again the basic selection criteria. This
sample is also flux density limited, with a threshold of $S_5>150$ mJy
(corresponding to the B3/VLA CSS sample selection for a spectral index
$\alpha_{0.4}^5 = 0.67$).

These selection criteria point to a population that is quite similar to the one
sampled by the B3/VLA CSS works: the median redshift is $z=1.085$, the radio
luminosity range is $10^{26}<P_{1.4}<10^{28}$ [W Hz$^{-1}$]. Indeed, the two
studies partly overlap, with a handful of objects found in both samples.

The sample has been studied in detail in a series of 5 papers, devoted to the
study of the morphology of the objects on different angular scales, using
MERLIN and VLA on large scale \citep{Kunert2002,Kunert-Bajraszewska2005},
MERLIN and EVN/VLBA for smaller sources \citep{Marecki2006}, and the VLBA for
the most compact objects
\citep{Kunert-Bajraszewska2006,Kunert-Bajraszewska2007}. In total, images have
been presented for 46 sources, including 11 MSOs.

One of the goals of this project was to understand whe\-ther low power CSS
sources were the progenitors of classic FR1 sources. No FR1 morphology was
revealed for sources with\-in the host galaxy, implying that pressure plays a
significant role. On the other hand, it has been proposed on the basis of the
morphology and nuclear powers that some sources in the sample could be
candidate restarted or fading sources. This includes a broad absorption line
(BAL) quasar, with asymmetric two-sided jet morphology.

To better clarify the evolutionary paths of all compact sources, a second
sample selected from the FIRST survey has also been presented in this workshop
\citep{Kunert-Bajraszewska2009}. This new sample does not overlap with the
previous one and it is focused on less bright sources (70 mJy $<S_{1.4}<$ 1.0
Jy) with steeper spectral index ($\alpha_{1.4}^5 > 0.7$), thus increasing the
incidence of low power and candidate fader compact sources.

\subsubsection{The GPS 1 Jy sample}

The two above samples have selection criteria that exclude sources with a
spectral peak above a few GHz. Indeed, a selection of CSS is relatively
straightforward to obtain, since the spectrum is optically thin in the GHz
domain. If we are interested in smaller and presumably younger radio sources,
we need to consider samples of GPS sources, whose selection is certainly a more
complex process. Mostly because of this, one of the main GPS sample available
is still the GPS 1 Jy sample presented by \citet{Stanghellini1998} and already
discussed at the Kerastari meeting \citep{Stanghellini2003}.

The GPS 1 Jy sample consists of 33 objects, selected with a spectral peak in
the range $0.4 < \nu_\mathrm{peak} < 6$ [GHz], a steep spectral index in the
optically thin part ($\alpha > 0.5$) and a 5 GHz flux density $S_5>1$ Jy.  As
we consider objects with higher spectral peak, the host start to become
contaminated by quasar, although $>50\%$ of them are still galaxies.

Among the most recent results derived from this sample, we can refer to the
follow up works on the nature of its members, carried out by
\citet{Stanghellini2005}, who looked at the presence of extended radio emission
or with VLBI observations. A search for molecular gas has also been performed,
but none was found \citep{O'Dea2005}.


\subsubsection{The Parkes 0.5 Jy GPS sample}

A southern counterpart to the GPS 1 Jy sample is the Parkes 0.5 Jy sample
\citep{Snellen2002}. Selected from multi-wavelength Parkes data, it is somewhat
deeper ($S_{2.7}>0.5$ Jy) and contains 49 sources.

An interesting feature of this sample is that it contains only sources
identified with galaxies; the host galaxies have all been identified
\citep{deVries2007}, and redshifts are measured for 80\% of sources, yielding a
range of $ 0.17<z<1.53$. Recent EVN observations of 15 sources show
mini-double-lobe radio structure \citep{Liu2007}, confirming that CSO
morphologies outnumber core-jets when only galaxies-hosted GPS are considered.

\subsubsection{Extreme GPS samples: High Frequency Peakers}

The problem of the contamination by core-jet variable sour\-ces becomes really
dramatic when extreme GPS sources are considered. A sample of so-called
High-Frequency Peaker (HFP, $\nu_\mathrm{peak}>5$ GHz) sources has been
selected by \citet{Dallacasa2000} in order to find objects with the smallest
$LS$ and therefore possibly the smallest age ($<500$ yrs): a cross correlation
of the Green Bank (GB) survey and the NVSS, followed up by multi-frequency VLA
observations, has yielded a sample of 55 genuine (i.e.\ not variable) HFPs.

A large fraction of objects are hosted by quasars (36/55).  Two frequency VLBA
observations have revealed a di\-cho\-tomy between galaxy hosted HFPs, which
have double or triple morphology, and QSO hosted HFPs, which tend to be
unresolved or core-jets \citep{Orienti2006}. Other clues on the nature of HFPs
in the sample have been obtained by studying HI absorption, spectral
variability, and polarization, and are discussed in the exhaustive review given
at this conference by \citet{Orienti2009}.


\subsection{Samples selected on morphology}

\subsubsection{The CORALZ sample}

The B3/VLA and the FIRST based samples were specifically designed to focus on
CSS sources selecting on spectral index. Since this can somehow bias the
results (e.g.\ excluding the most compact sources), \citet{Snellen2004} have
selected a sample of Compact Radio sources at Low Redshift (CORALZ) without any
spectral index constraint. The sample is obtained from cross-correlation of the
FIRST with the APM/POSS and WENSS regions, selecting sources with a
$S_\mathrm{peak} >100$ mJy, a $LAS <2\arcmin$, and a redshift $z<0.2$.

The sample consists of 28 objects, and it is 95\% complete in the range
$0.005<z<0.16$ (17 objects). Interestingly, an {\it a posteriori} look at the
spectra of these sources confirms that they are essentially all CSS (10) or GPS
(6). From a morphological point of view, EVN, MERLIN, and Global VLBI images
reveal that about a half of the sources in the sample are CSOs
\citep{deVries2009}. Since the sources are all relatively nearby, an accurate
determination of proper motion in the components should actually be feasible
within relatively short time scales.

Another remarkable consequence of the low distance of the objects in the sample
is that they have quite low radio luminosity, namely in the range $ 22.96 <
\mathrm{Log} P_5 < 25.25$ [W Hz${-1}$]. Finally, it should also be easier to
follow up the sources in other wavebands, such as the X-rays or the CO lines
\citep[see e.g.][]{Mack2009}.

\subsubsection{The COINS and VIPS samples}

If one neglects the spectral classification, an interesting way to select
candidate young sources is to directly look at their parsec scale
structure. This is ideally done by looking at the images available thanks to
large VLBI surveys. A successful example of such an approach is the sample of
CSOs Observed In the Northern Sky \citep[COINS,][]{Peck2000}. It contains 52
candidates, selected from large VLBI surveys with $ S_5>100$ mJy and nearly
equal double structure or core with emission on either sides.

The selection criteria themselves makes VLBI follow ups relatively more
straightforward. Multi-frequency, multi-epoch VLBI observations have confirmed
the CSO nature of at least 17 candidates and allowed the measurement of hot
spot advance velocities (or limits). The works on the COINS sample by
\citet{Gugliucci2005,Gugliucci2007} have indeed provided the largest amount of
new age estimates (3 sources, ranging between $20 \pm 4$ to $3000 \pm 1490$, in
years) and limits (9 sources, in the range 280--2220 yr) since the Kerastari
meeting (see also Sect. \ref{ages}).


A similar approach has been applied more recently to the large dataset provided
by the VLBA Imaging and Polarimetry Survey (VIPS), which contains 1127 sources
imaged at 5 GHz \citep{Helmboldt2007}. A total of 103 CSO candidates have been
selected from this survey and are currently under review for confirmation
\citep{Tremblay2009}.

\subsection{Samples: summary}


As discussed in the previous sections, the selection of samples of
intrinsically small sources can be done on the basis of the spectral properties
or on the morphology. Both approaches have advantages and disadvantages. On one
hand, it is relatively straightforward to obtain a starting list simply by
cross-correlating catalogues at different frequencies. Of course, things become
more and more complex when one wants to consider sources with higher
$\nu_\mathrm{peak}$: high frequency catalogues are less numerous and extended;
contamination by core-jet sources and flaring blazars can become
significant. At some point, the information on morphology becomes necessary.

On the other hand, starting from the information on morphology can provide from
the very beginning a lot of useful information. The distinction between truly
compact sources and beamed core-jets is somewhat simpler, and the discussion of
advance motions can often be started with just one additional set of
observations.  However, the availability of the initial data is not as large as
in the case of spectral selection and, moreover, these master lists can be
severely biased. For example, there are large intervals of spatial frequencies
that are not well sampled by surveys with present instruments. This makes a
discussion of the statistical significance of the obtained results certainly
more difficult. See also on this topic the relevant work done on MSO by
\citet{Augusto2006,Augusto2009}.

In any case, both spectrum and morphology are essential pieces of
information. Whatever the starting approach, there is always an interaction
between the two methods, whose complementarity is clear. Finally, it is also
worth mentioning an entirely different methodology, such as the one based on
the low polarization criterion \citep*[see e.g.][]{Cassaro2009}.


\section{Statistics} \label{s.stats}

As we showed in the previous section, there has been clearly a remarkable
amount of work on the topics of selecting and studying samples of compact
sources. Numbers are therefore getting much larger than what was available in
2002. As for CSS sources, the master lists of the B3/VLA and the two FIRST
based samples sum up to a total of 191 sources (although in the end some of
them were rejected/duplicated). The radio luminosity range is also being pushed
to lower values, in the effort to include also low power CSS with FR1
morphologies.  The B3/VLA and the brightest FIRST sample extend over the
luminosity range $10^{25}<P_{0.4}<10^{29}$ [W Hz$^{-1}$], while the new FIRST
based sample is eventually trying to get below the 10$^{25}$ W Hz$^{-1}$
threshold.

Indeed, \citet{Giroletti2005} have presented at least a couple of compact
sources that are FR1-like both in morphology and radio luminosity: 0258+35
($z=0.016$, $P=10^{24.4}$ W Hz$^{-1}$, $LS = 1.5$ kpc), and 1855+37 ($z=
0.055$, $P=10^{24.6}$ W Hz$^{-1}$, $LS = 7$ kpc). However, these sources are
missed in present samples, either because of a spectral index slightly flatter
than $\alpha=0.5$ (0258+35) or a low flux density ($S_{0.4} < 0.8$ Jy,
1855+35).

Another boundary to cross is on the size, with an interest on the more elusive
MSO. \citet{Augusto2009} reports an interesting search for such objects,
considering 157 candidates.

Numbers are also getting large if we consider sources with higher
$\nu_\mathrm{peak}$. \citet{Labiano2007} have compiled a master list of 172 GPS
and HFPs. Here, the main problem is still discussing the optical counterpart:
59 sources are hosted by QSOs, 82 by galaxies, and 31 are still empty
fields. Redshifts are available for 108 sources (63\%), covering the range
$0.008<z<3.77$, with a median value of $z=0.76$. The flux densities are in the
range 47 mJy$ < S < $6.5 Jy (median 0.77 Jy, mean 1 Jy), although the different
spectral properties make this values not entirely homogeneous.

As anticipated by \citet{Woltjer2003}, we are starting to capitalize on such
large numbers in order to get information about volume densities and
evolution. For example, \citet{Tinti2006} considered a sample of 111 GPS and
found that the observed redshift and peak frequency distributions are in
agreement with simple luminosity evolution of individual sources. A decrease of
the emitted power and of the peak luminosity with source age or with decreasing
peak frequency was also required. Completion of the information on the host
properties, including redshift measurement, and a less ambiguous selection of
the GPS samples could help to converge on this model.

\subsection{Ages} \label{ages}

\begin{table*}
 \centering
\caption{Hot spot advance velocity and kinematic age estimates. Unless otherwise noted, measurements are from hot spot to hot spot.}
\label{t.ages}
\begin{tabular}{lcrrrc}\hline
Source   & $z$    & Size & Velocity & Age & Reference \\
         &        & (pc) & ($c$)    & (yr) & \\
\hline
0035+227 & 0.096 &  21.8 & 0.15 & 450  & 1 \\
4C31.04	(0116+319) & 0.059 & 70.1  & 0.45 & 501  & 2 \\
0108+388 & 0.669 &  22.7 & 0.18 & 404  & 1 \\
J0204+0903 & n/a & 18.3* & 0.07* & 240 & 3 \\ 
J0427+4133 & n/a & 1.3*  & 0.06* & 20  & 3 \\
0710+439 & 0.518 &  87.7 & 0.30 & 932  & 4 \\
1031+567 & 0.460 & 109.0 & 0.19 & 1836 & 4 \\
1245+676 & 0.107 &   9.6 & 0.16 & 188  & 4 \\
OQ208 (1404+286)  & 0.077 &   7.0 & 0.10 & 219  & 5 \\
CTD 93 (1607+268) & 0.473 & 240 & 0.34 & 2200 & 6 \\
1718-649 & 0.014 & 2.0	 & 0.07 & 91   & 1 \\
J1826+1831 & n/a & 41.9* & 0.015* & 2600 & 7 \\
1843+356 & 0.763 &  22.3 & 0.39 & 180  & 4 \\
1934-638 & 0.183 &  85.1 & 0.17 & 1603 & 1 \\
1943+546 & 0.263 & 107.1 & 0.26 & 1308 & 4 \\
1946+708 & 0.101 & 39.4  & 0.10 & 1261 & 1 \\
2021+614 & 0.227 &  16.1 & 0.14 & 368  & 4 \\
2352+495 & 0.238 & 117.3 & 0.12 & 3003 & 4 \\
%
%
%
\hline
\end{tabular}

(*): Size in mas, velocity in mas yr$^{-1}$. Motion is measured from hot spot
to core.

References: 1. This work; 2. \citet{Giroletti2003}; 3. \citet{Gugliucci2005};
4. \citet{Polatidis2003} and references therein; 5. \citet{Luo2007};
6. \citet{Nagai2006}; 7. \citet{Gugliucci2007}.
\end{table*}

Surely, we are getting also some progress on the evolution of individual
sources. \citet{Polatidis2003} had summarized available kinematic age studies,
reporting 10 estimates and 3 lower limits. Observations in recent years have
contributed to increase the number of measurements and in some cases to refine
previous figures. In Table \ref{t.ages}, we report the current available
estimates from the literature or from new data. Data are typically measured
from hot spot to hot spot.

The number of measurements has almost doubled.  In total, there are now 18
estimates, ranging from 20 to 3000 years (mean age 967 yrs, median 435
yrs). Sources span a range in linear size between 2 and $\sim120$ parsecs.  The
hot spot velocities are still consistent with the values reported by
\citet{Polatidis2003}. The current average velocity is $v_\mathrm{sep} = (0.19
\pm 0.11)h^{-1}c$. A number of lower limits have also been found
\citep[e.g. by][]{Gugliucci2005}, and non-radial motions seem also to be
present in a few sources \citep{Stanghellini2009}.

Whether all sources actually show hot spot motion is still to be solved. Limits
on motions have been found in several cases
\citep[e.g. by][]{Gugliucci2005}. Moreover, non-radial motions seem also to be
present in a few sources \citep{Stanghellini2009} and it has been reported at
least one case of inward moving hot spots \citep{Tremblay2008}. It is
encouraging to report several new works presenting VLBI images of large numbers
of GPS and CSOs, e.g.\ \citet[][19 GPS sources]{Liu2007}, \citet[][103 CSO
  candidates]{Tremblay2009}, \citet[][for the CORALZ sample]{deVries2009}. One
can then expect an increase of advance motion measurements in the next few
years.

\section{Looking forward to the next meeting...} \label{s.future}

Our community has certainly made a great advance in the understanding of the
physics and evolution of radio sources thanks to the samples discussed
here. Upcoming new instrumentation can be used to take further important steps.
The need to consider low power sources seems to be already quite clear; the
increased sensitivity of e-VLA and e-MERLIN and the use of phase referenced
VLBI observations will certainly help this cause. e-VLA and e-MERLIN will also
be important to obtain images on the intermediate scales somehow neglected so
far. LOFAR will also offer a great opportunity to assess the fraction of young
sources that show evidence of remnants of previous stages of activity, while
ALMA could discover the youngest radio sources with the highest
$\nu_\mathrm{peak}$.

The physics of the radio sources will also be probed thanks to high-energy
astronomy, and in particular by the recently launched {\it Fermi Gamma-ray
  Space Telescope} (formerly GLAST) which with its sensitivity and monitoring
capability will provide statistically significant results. In any case, the
search for outliers and the monitoring will remain an important task.  It is
also likely that constraints on spectral index will have to be released in
order to obtain a coherent view of various populations.

Space VLBI is also entering a new era. In particular, VSOP2 will be launched in
2012 and it could be a fantastic machine for hot spot advance measurements. The
angular resolution will be as high as 38 $\mu$as at 43 GHz and will
significantly shorten the time scales for the detection of motions in the most
compact sources. A pre-launch survey could be necessary to find the best
candidates - and in particular to assess the numbers of sources that can be
most successfully studied at 8, 22, and 43 GHz.




\end{document}